\documentclass[letterpaper]{article} 
\usepackage{aaai25}  
\usepackage{times}  
\usepackage{helvet}  
\usepackage{courier}  
\usepackage[hyphens]{url}  
\usepackage{graphicx} 
\usepackage{natbib}  
\usepackage{caption} 
\usepackage{algorithm}
\usepackage{algorithmic}
\usepackage{amsmath}
\usepackage{newfloat}
\usepackage{listings}
\DeclareCaptionStyle{ruled}{labelfont=normalfont,labelsep=colon,strut=off} 
\floatstyle{ruled}
\newfloat{listing}{tb}{lst}{}
\floatname{listing}{Listing}
\usepackage{amsfonts,amssymb}
\usepackage{booktabs}
\usepackage{adjustbox}
\usepackage{multirow}
\usepackage{makecell}
\usepackage{diagbox}
\usepackage{subfigure}
\usepackage{marvosym}

\title{CoRA: Collaborative Information Perception by Large Language Model's Weights for Recommendation}
\author {
    Yuting Liu\textsuperscript{\rm 1},
    Jinghao Zhang\textsuperscript{\rm 2},
    Yizhou Dang\textsuperscript{\rm 1},
    Yuliang Liang\textsuperscript{\rm 1},\\
    Qiang Liu\textsuperscript{\rm 2},
    Guibing Guo\textsuperscript{\rm 1}$^{(\textrm{\Letter})}$,
    Jianzhe Zhao\textsuperscript{\rm 1},
    Xingwei Wang\textsuperscript{\rm 1}
}
\affiliations {
    \textsuperscript{\rm 1}Software College, Northeastern University\\
    \textsuperscript{\rm 2}New Laboratory of Pattern Recognition (NLPR), Institute of Automation, Chinese Academy of Sciences,\\
    \{liuyuting, yizhoudang, liangyuliang\}@stumail.neu.edu.cn, jinghao.zhang@cripac.ia.ac.cn, qiang.liu@nlpr.ia.ac.cn,\{guogb, zhaojz\}@swc.neu.edu.cn, wangxw@mail.neu.edu.cn
}

\usepackage{bibentry}

\begin{document}

\maketitle

\begin{abstract}
Involving collaborative information in Large Language Models (LLMs) is a promising technique for adapting LLMs for recommendation. 
Existing methods achieve this by concatenating collaborative features with text tokens into a unified sequence input and then fine-tuning to align these features with LLM's input space.
%
Although effective, in this work, we identify two limitations when adapting LLMs to recommendation tasks, which hinder the integration of general knowledge and collaborative information, resulting in sub-optimal recommendation performance. (1) Fine-tuning LLM with recommendation data can undermine its inherent world knowledge and fundamental competencies, which are crucial for interpreting and inferring recommendation text. (2) Incorporating collaborative features into textual prompts disrupts the semantics of the original prompts, preventing LLM from generating appropriate outputs.
In this paper, we propose a new paradigm, \textbf{Co}llaborative \textbf{Lo}RA (CoRA), with a collaborative query generator. Rather than input space alignment, this method aligns collaborative information with LLM's parameter space, representing them as incremental weights to update LLM's output. This way, LLM perceives collaborative information without altering its general knowledge and text inference capabilities. 
Specifically, we employ a collaborative filtering model to extract user and item embeddings and inject them into a set number of learnable queries. We then convert collaborative queries into collaborative weights with low-rank properties and merge the collaborative weights into LLM's weights, enabling LLM to perceive the collaborative signals and generate personalized recommendations without fine-tuning or extra collaborative tokens in prompts. 
Extensive experiments confirm that CoRA effectively integrates collaborative information into LLM, enhancing recommendation performance.
\end{abstract}

\section{Introduction}

Large language models (LLMs) have showcased remarkable performance in a wide range of natural language processing tasks and demonstrated excellent generalization capabilities, offering a promising solution to real-world problems. 
To leverage the extensive competencies of LLMs, an increasing number of studies are investigating ways to frame recommendation problems in natural language, allowing LLMs to tackle these queries (LLMRec)~\cite{Li2023TextIA,Zhang2024StealthyAO,Dai2023UncoveringCC}. Collaborative information, which delineates the co-occurrence patterns in user-item interactions~\cite{Textlike}, is critical for the success of conventional collaborative filtering methods. Regardless, in LLM-based recommendation approaches, collaborative information cannot be directly interpreted as text. LLMs' ineptitude in perceiving this information hampers their performance compared to traditional recommendation models. Consequently, enabling LLMs to comprehend and utilize collaborative information presents a significant challenge.

\begin{figure}
    \centering
    \includegraphics[width=\linewidth]{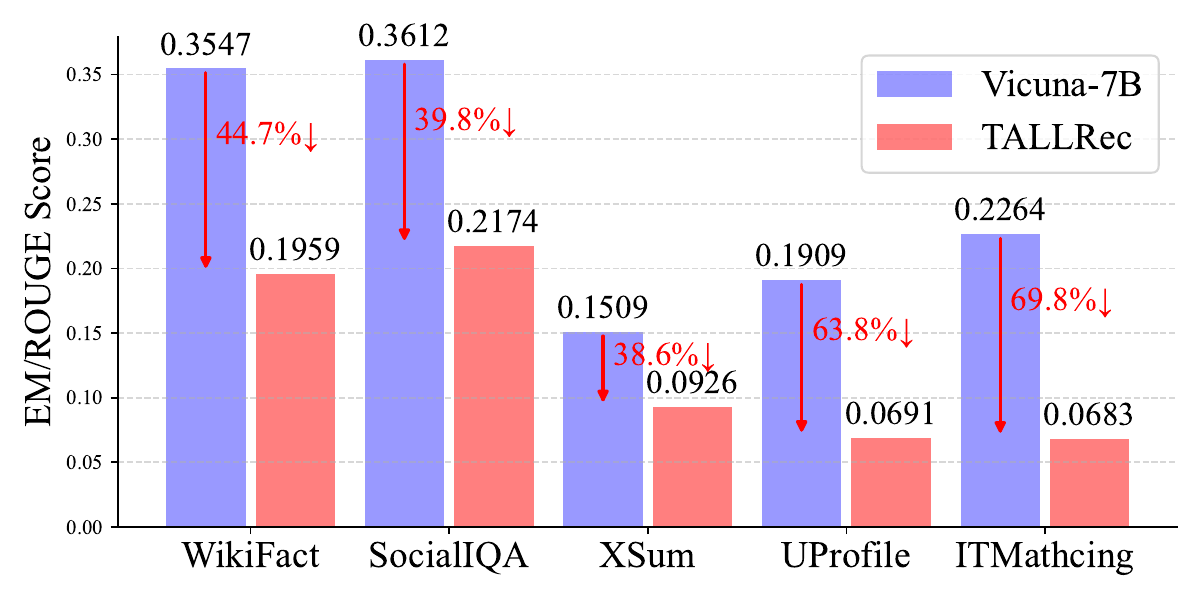}
    \caption{The performance of Vicuna-7B before and after fine-tuning on Amazon-Book using the prompt in TALLRec. The EM/ROUGE-L scores of generated answers on datasets represent various general and recommendation abilities.}
    \label{fig:knowledge}
\end{figure}

Some approaches are beginning to focus on this issue~\cite{Li2023CTRLCT,TOIS23-PEPLER}. For instance, Prompt4NR~\cite{prompt4nr} makes the first trial of \textit{prompt learning} paradigm to convert the News Recommendation task as a language prediction task, where user and item data are wrapped into textual prompts. CoLLM~\cite{CoLLM} captures collaborative information through an external traditional model and maps it to the input space of LLM, forming user and item embeddings as unique tokens for LLM usage. BinLLM~\cite{Textlike} converts collaborative embeddings into binary value sequences that LLMs can operate on directly, facilitating LLMs' direct usage of collaborative information in text-like format. LlaRA~\cite{llara} uses a hybrid prompting method that integrates ID embeddings with textual features.

Despite effectiveness, these techniques fail to preserve the LLM's inherent comprehension and inference for universal and recommendation information when adapting LLMs to recommendation. Our experiments reveal that LLM's capabilities have been weakened from two perspectives, leading to sub-optimal recommendation performance.

(1) Researchers have discovered that supervised fine-tuning on instructions they have never seen will encourage them to produce hallucinations~\cite{TheDawnAfterTheDark}. We evaluate LLM's interpreting and inferring capabilities for general and recommendation textual descriptions.
For general knowledge, we use Vicuna-7B\cite{vicuna2023} to generate answers on \emph{WikiFact}~\cite{wikifact}, \emph{SocialIQA}~\cite{socialiqa}, and \emph{XSum}~\cite{XSum} datasets. For recommendation knowledge, we constructed subsets for user profiling (UProfile) and item title matching (ITMatching) based on the \emph{Amazon-Book}~\cite{Amazon} to evaluate LLM's comprehension of user and item textual description\footnote{The evaluation details are provided in the Appendix.}.
As shown in Fig.~\ref{fig:knowledge}, the LLM's performance on all five datasets significantly decreased after fine-tuning on \textit{Amazon-Book} using the prompt in TALLRec~\cite{TALLRec}, indicating that its capacities for utilizing, reasoning, summarizing, profiling, and understanding recommendation texts have been compromised during fine-tuning. 
 

\begin{figure}
    \centering
    \includegraphics[width=\linewidth]{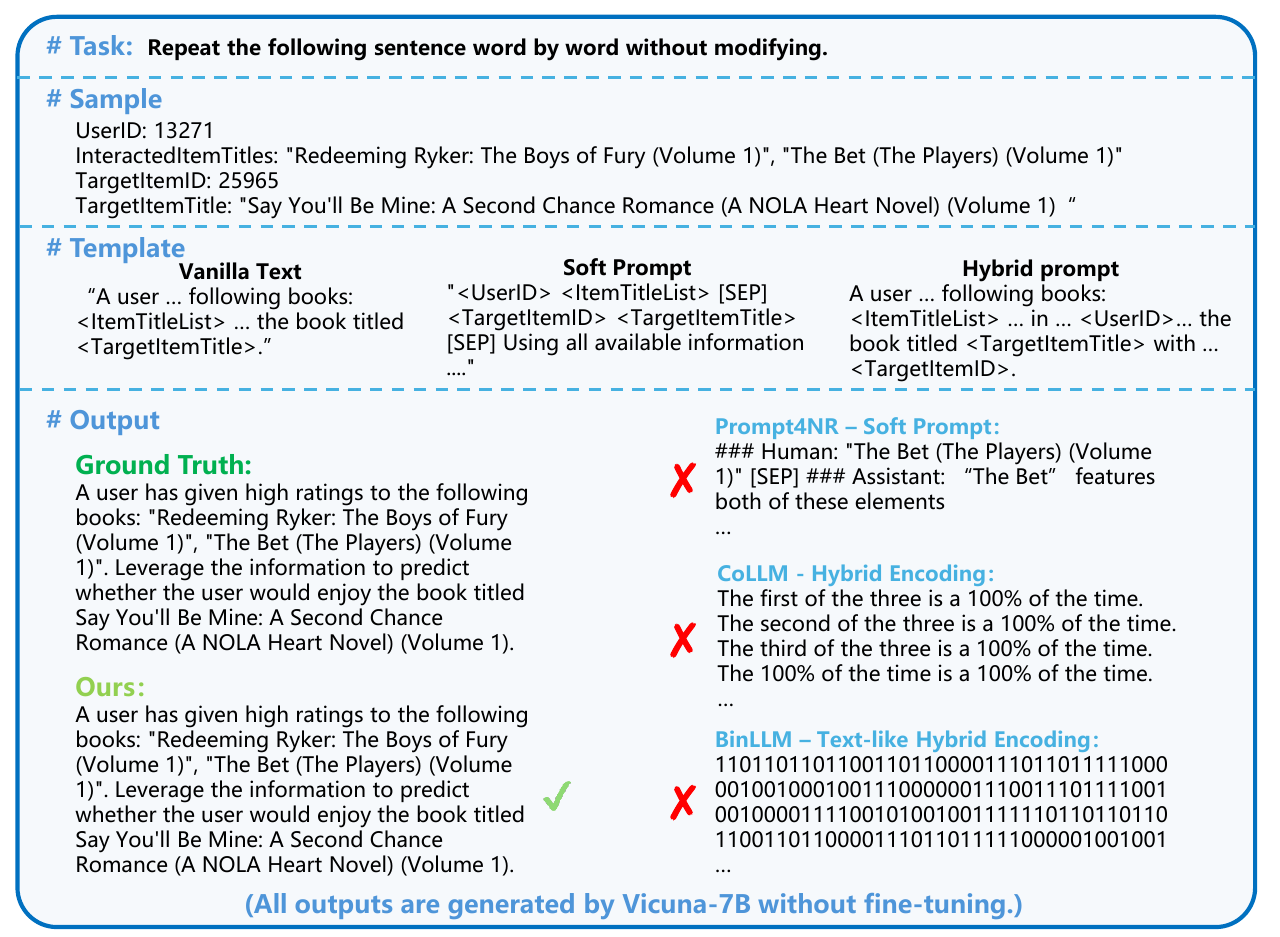}
    \caption{Collaborative features interfering with LLM's understanding of textual prompts. We use pre-trained Vicuna-7B as the ground truth. Our method avoids this interference.}
    \label{fig:example}
\end{figure}

(2) Existing methods endeavor to align collaborative information with the input space of LLMs by embedding collaborative features of users and items into textual prompts. Nevertheless, this practice can disrupt the LLM's comprehenson of the original text semantics. For instance, as shown in Fig.~\ref{fig:example}, while we expect Vicuna-7B without fine-tuning to repeat user and item description, the revised prompts fail to guide the LLM to generate the correct responses. Overall, existing studies limit general knowledge and collaborative information integration when adapting LLM to recommendation tasks.



To address this issue, we propose aligning collaborative features with the LLM's parameter space by transforming them into plug-in weights to update LLM's output. Unlike input space alignment, this parameter space alignment approach enables general LLM to directly leverage collaborative information as an adaption of each user-item pair for recommendation tasks, eliminating the need for fine-tuning or modifying the text prompts.

Specifically, the prediction for each user-item pair is divided into two parts: collaborative weights generation and text inference with collaborative weights. First, we utilize a pre-trained collaborative filtering model to extract user and item embeddings, which are processed by a collaborative query generator. The generator receives collaborative information and injects it into learnable query embeddings. Inspired by VLoRA~\cite{VLoRA} in computer vision, we transform collaborative queries into collaborative perceptual weights similar to ``low-rank adaption" (LoRA~\cite{lora}). We incorporate the collaborative weights into the LLM's pre-trained weights to endow it with the capability to perceive collaborative information between the user and the item, allowing it to generate personalized predictions for each user-item pair with general text prompts. Importantly, our technique does not alter the LLM's understanding of the original text, as it does not introduce new tokens. Instead, it updates LLM's output with incremental weights. For tasks that do not involve collaborative information, our method seamlessly switches to utilizing the frozen backbone to infer textual prompts. Therefore, our procedure inherently addresses the aforementioned problems.
To sum up, our contributions are summarized as follows:
\begin{itemize}
    \item We explore the issues present when integrating textual prompts and collaborative features in LLM's input space. To resolve these matters, we propose to equip LLMs with collaborative perception capability by merging collaborative weights into LLM's pre-trained weights. It can be seen as \textbf{Co}llaborative Lo\textbf{RA} (CoRA), facilitating the cooperation of text and collaborative information. To the best of our knowledge, we are the first to endeavor collaborative perception in the parameter space of LLMs.
    \item Following the CoRA method, we propose a collaborative query generator that injects collaborative features extracted from conventional recommenders into learnable query embeddings and then transforms them into LLM weights space with a low-rank property.
    \item Experimental results demonstrate the effectiveness of our approach, showing significant improvements over state-of-the-art LLMRec methods and traditional collaborative filtering methods on real-world datasets.
\end{itemize}

\section{Related Work}

\subsection{Collaborative Filtering Models}

Collaborative information is essential in the existing recommendation literature\cite{IDvsModal,idsf}. In traditional personalized recommendation, collaborative filtering (CF) models are prevalent, leveraging collaborative information of users and items to generate predictions~\cite{bpr,Guo2015TrustSVDCF}. In these models, users and items are represented as latent factors fed into neural networks to model their interactions~\cite{lightgcn,Tang2018PersonalizedTS,He2017NeuralCF}. These studies achieved remarkable success in academia and industry, inspiring further exploration into collaborative information perception for LLMRec.

\subsection{LLM for Recommendation}

Given the impressive capabilities exhibited by LLMs, there is an increasing focus on exploring their potential applications in recommender systems. Techniques in this domain involve translating recommendation tasks into natural language tasks and adapting LLMs to generate recommendation results directly. These generative approaches can be divided into two paradigms based on whether parameters are tuned: non-tuning and tuning paradigms. The non-tuning paradigm assumes LLMs already possess the recommendation abilities and attempts to leverage their strong zero/few-shot abilities by introducing specific prompts~\cite{Liu2023IsCA,Dai2023UncoveringCC,Mysore2023LargeLM, Wang2023RethinkingTE,Hou2023LargeLM}. In contrast, the tuning paradigm uses prompt learning or instruction tuning~\cite{Kang2023DoLU,Wang2022TowardsUC,cui2022M6Rec} to enhance LLM's recommendation capacities.

A new trend is surfacing to include collaborative information in LLMs. Some studies focus on discovering user and item encoding methods to introduce new tokens through vocabulary expansion\cite{Zheng2023AdaptingLL,Hua2023HowTI,Rajput2023RecommenderSW,Zheng2024HarnessingLL}. Others explore extracting collaborative information using latent factor models and aligning it with the input space of LLMs~\cite{CoLLM,Textlike,llara,prompt4nr,Li2023E4SRecAE}. While these methods exhibit strong performance, they constrain specific general capabilities of LLMs. Moreover, due to the hybrid prompting methods, collaborative features can disrupt original textual semantics. Motivated by VLoRA~\cite{VLoRA}, which enables LLMs to perceive visual features by representing visual information as model weights, we propose CoRA, which generates low-rank collaborative weights like LoRA and injects them into LLM's weights without extra inference overhead.

\section{Preliminaries}

In this section, we introduce the problem definition, the basic concepts, and the notations in this paper.

\subsection{Problem Definition}

We represent the interaction dataset as $\mathcal{D}=\{(u,i,y)|u\in\mathcal{U},i\in\mathcal{I}\}$ where $\mathcal{U}$ and $\mathcal{I}$ denote the set of users and items, respectively, with $y_{ui}=\{0,1\}$ indicating the interaction label. For an item $i$, there is a unique identifier and textual description $t_i$. For a user $u$, we construct the textual description from its historical interactions $t_u=\{t_i|i\in\mathcal{I}_u\}$. In this study, we aim to enable LLMs to leverage both collaborative embeddings $e_u, e_i$ and textual descriptions $t_u, t_i$ to predict whether a user $u$ will enjoy an item $i$ (i.e. $y_{ui}$).

\subsection{Collaborative Filtering Models}

To gather collaborative information, we are looking into utilizing CF methods, which represent users and items by latent factors (a.k.a. embeddings). User and item embeddings are variously calculated to capture collaborating relations precisely. The formulation of encoding a sample $(u,i,y)\in\mathcal{D}$ can be written as follows:
\begin{equation}
    \mathbf{e}_u=f_{\psi}(u;\mathcal{D});\ \mathbf{e}_i=f_{\psi}(i;\mathcal{D}),
\end{equation}
where $\mathbf{e}_u,\mathbf{e}_i\in\mathbb{R}^{d_c}$ denote the user's and item's representations with dimension $d_c$, $f_{\psi}(\cdot)$ denotes the encoding process, and $\psi$ represents model parameters. The user and item embeddings are then fed into the interaction prediction.

\subsection{Large Language Model}
\label{sec:llm}

\begin{figure}
    \centering
    \includegraphics[width=\linewidth]{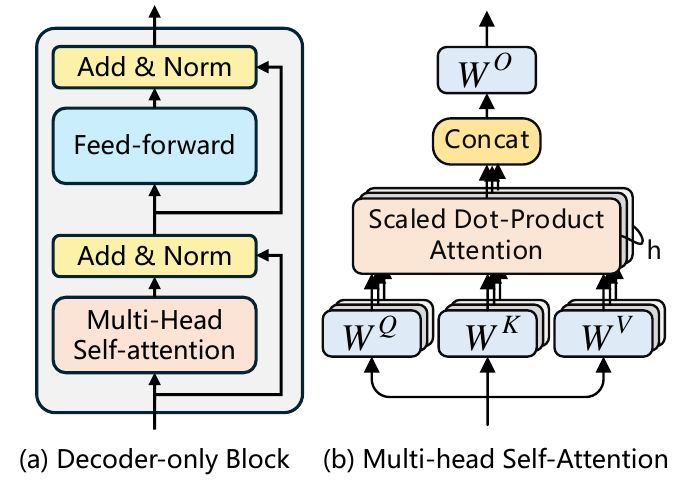}
    \caption{(a) Architecture of the LLM's Decoder Block. (b) Details of the multi-head self-attention module.}
    \label{fig:llm}
\end{figure}

LLMs refer to a class of language models equipped with billions of parameters. Due to its superiority in performance and training efficiency in generative tasks, LLMs with decoder-only architecture have become mainstream. As depicted in Fig.~\ref{fig:llm}, their decoder block contains a multi-head self-attention module, two add-and-norm modules, and a feed-forward network~\cite{transformer}.

\textbf{Multi-Head Self-Attention} module consists of four types of weights: $W^Q$, $W^K$, $W^V$, and $W^O\in\mathbb{R}^{d_\text{model}\times d_\text{head}}$, where $d_\text{model}$ and $d_\text{head}$ are dimensions of the input embeddings and attention heads, and $d_\text{head}$=$d_\text{model}/N_\text{head}$. For an input sequence with $L$ tokens $X\in\mathbb{R}^{L\times d_\text{model}}$, the calculation of the multi-head self-attention layer is as follows:
\begin{align}
    \mbox{MultiHead}(Q,K,V)&=\mbox{Concat}(\mbox{head}_1,\dots,\mbox{head}_{N_\text{head}})W^O \nonumber\\
        \mbox{where}\ \mbox{head}_i&=\mbox{Attention}(Q_i,K_i,V_i)\\
        &=\mbox{softmax}(\frac{XW^Q_iXW^{K^\top}_i}{\sqrt{d_\text{head}}})XW^V_i.\nonumber
\end{align}

\begin{figure*}
    \centering
    \includegraphics[width=0.8\linewidth]{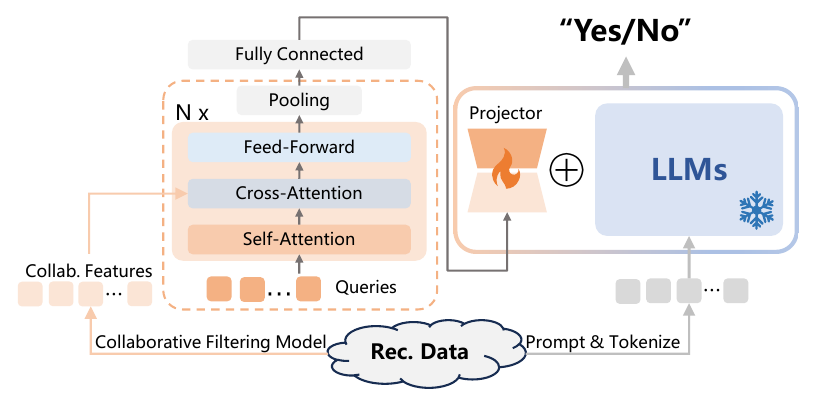}
    \caption{Model architecture overview of our CoRA. The left path extracts user and item embeddings using a CF model and generates collaborative queries. The right path fills the text fields in the prompt template, introducing textual descriptions for inference. Finally, the collaborative queries are projected into the LLM's parameter space and merged into the LLM's weights, enabling the LLM to perceive collaborative information without any fine-tuning or extra tokens in textual prompts.}
    \label{fig:model1}
\end{figure*}

\textbf{Feed-forward Network} is an MLP with two linear layers and one non-linear activation function as follows:
\begin{equation}
    \mbox{FFN}(X)=\sigma(XW_\text{up})W_\text{down},
\end{equation}
where $W_\text{up}$ and $W_\text{down}$ are the weights of linear layers, and $\sigma$ is an activation function. Overall, there are six types of linear weights in a decoder block: $W^Q$, $W^K$, $W^V$, $W^O$ in the multi-head self-attention module, and $W_\text{up}$, $W_\text{down}$ in the feed-forward module.

\section{Methodology}

In this section, we introduce our collaborative weights generator and CoRA method, which effectively integrates collaborative information into LLMs in the parameter space. The overall framework is shown in Fig.~\ref{fig:model1}. First, we explain how to obtain collaborative queries from the pre-trained user and item embeddings through a collaborative query generator. Then, we elaborate on how to equip LLM with collaborative perception capability with the help of generated collaborative queries, followed by a description of the prediction and training strategy.

\subsection{Generating Collaborative Queries}


For a user-item pair $(u,i)$, we first obtain user and item embeddings $\mathbf{e}_u,\mathbf{e}_i\in\mathbb{R}^{d_c}$ from a pre-trained CF model and concatenate them as $[\mathbf{e}_u,\mathbf{e}_i]$. Then, we fed them into a collaborative query generator to obtain user-item-specific collaborative queries, which bridge the collaborative information and the LLM's parameter space. Similar to the Q-Former in BLIP-2~\cite{blip2} and the perceptual weights generator in VLoRA~\cite{VLoRA}, we initialize $k$ learnable query embeddings and input them into $N$ decoder blocks with cross-attention modules, to absorb collaborative information from different $k$ semantic sub-spaces. 

Specifically, as shown in Fig.~\ref{fig:model1}, $k$ query embeddings first interact with each other in the self-attention module, capturing the underlying relationship between different representation sub-spaces. Then, these $k$ query embeddings perceive and comprehend collaborative information in pre-trained user and item embeddings with the help of the cross-attention module. Finally, the $k$ query embeddings are transformed into $k$ deep collaborative features after passing through the feed-forward network. Afterward, we adopt a pooling operation to aggregate collaborative information from $k$ sub-spaces, obtaining the eventual collaborative information-aware query embedding $\mathbf{q}_c\in\mathbb{R}^{2d_c}$, where $d_c$ is the dimension of pre-trained user/item embeddings. In BLIP-2, the output query embedding is linearly projected into the same dimension as the text embedding of the LLM. However, to avoid the collision between textual prompts and collaborative information, we utilize it to generate an incremental weight to guide LLM in recognizing collaborative information and generating appropriate results. 

\subsection{Collaborative Perception in LLM}

Existing LLMRec methods project user and item embeddings into the LLM's input space to integrate collaborative and textual information. These approaches interfere with the LLM's comprehension of the original textual prompts using its general capabilities, resulting in sub-optimal performance. Consequently, we suggest avoiding the incorporation of user/item embeddings and textual prompts. Inspired by the success of VLoRA~\cite{VLoRA} in computer vision, we address this issue by treating the collaborative information as incremental weights of pre-trained LLMs. 

By establishing a bridge between collaborative information and the LLM's parameter space through query embedding $\mathbf{q}_c$, our objective is to generate collaborative weights $W_c$ that can be seamlessly integrated into the pre-trained LLM weights. As we mentioned in the preliminaries, the shape of LLM weights should be $d_\text{model}\times d_\text{model}$. With transforming $\mathbf{q}_c\in\mathbb{R}^{2d_c}$ into $\Delta W\in\mathbb{R}^{d_\text{model}\times d_\text{model}}$ directly, we will introduce a transformation matrix $W\in\mathbb{R}^{2d_c\times d_\text{model}\cdot d_\text{model}}$, whose parameter cost is unacceptable. Accordingly, a low-rank computation is considered in this process to reduce the computation costs. Inspired by LoRA~\cite{lora}, we utilize a Fully Connected layer to map the output query embedding $\mathbf{q}_c$ into LLM's parameter space as $\Delta W_A\in\mathbb{R}^{d_\text{model}\times r}$ in LoRA, and ascend the dimension with a final linear projector $W_\text{proj}\in\mathbb{R}^{r\times d_\text{model}}$, which is equivalent to $\Delta W_B$ in LoRA. Eventually, the user-item-specific collaborative weight is calculated by multiplying $\Delta W_A$ and $\Delta W_B$, which is formulated as:

\begin{equation}
    \begin{aligned}
        W_c&=\Delta W_A\Delta W_B\\
        &=\mbox{R}(\mathbf{q}_cW_\text{FC})W_\text{proj},
    \end{aligned}
\end{equation}
where $\mbox{R}(\cdot)$ represent a \textit{reshape} operator, which reshape the product of $\mathbf{q}_c$ and $W_\text{FC}$ from $\mathbb{R}^{d_\text{model}\cdot r}$ into $\mathbb{R}^{d_\text{model}\times r}$.


Generally, for an input sample consisting of a user, an item, and their textual descriptions, we extract their collaborative features through a pre-trained CF model and inject the features into a set number of learnable queries with the cross-attention module. Then, we map the collaborative queries into LLM's parameter space and transform them into incremental LLM weights with low-rank properties. Finally, the collaborative weights can be directly merged into pre-trained LLM weights:
\begin{equation}
\begin{aligned}
    \hat{W}=W+W_c=W+\mbox{R}(\mathbf{q}_cW_\text{FC})W_\text{proj},
\end{aligned}
\end{equation}
where $W\in\mathbb{R}^{d_\text{model}\times d_\text{model}}$ denotes the overall weights of LLM. By absorbing collaborative weights, the LLM inherently perceives collaborative information without altering the LLM’s understanding of the original textual prompts. 


\subsection{LLM prediction and Training Method}

\begin{table}
\centering
    \begin{tabular}{p{8cm}}  
    \toprule
        \#Question: A user has given high ratings to the following movies: $\langle$ItemTitleList$\rangle$. Leverage the information to predict whether the user would enjoy the movie titled $\langle$TargetItemTitle$\rangle$? Answer with ``Yes" or ``No". \textbackslash n\#Answer:\\
    \bottomrule
    \end{tabular}
    \caption{Example of the used prompt template, using the same format as TALLRec~\cite{TALLRec}.}
    \label{tab:prompt}
\end{table}

Once LLM is equipped with collaborative perceptual capabilities, it can predict recommendations without additional fine-tuning. For an input sample $s=(u, i, t_u, t_i)$, the prediction generation can be formulated as:
\begin{equation}
    \begin{aligned}
        \hat{y}&=\mbox{LLM}(s)=\mbox{LLM}_{W+\Delta W}(p) \\
        &=\mbox{LLM}_{W+g_{\Theta}([f_{\psi}(u),f_{\psi}(i)])}(p)
    \end{aligned}
\end{equation}
where $u,i$, and $t_u,t_i$ represent the identifier and textual description of the user and item, respectively. $p$ is the prompt constructed with textual information $t_u$ and $t_i$ as shown in Tab.~\ref{tab:prompt}. $g_{\Theta}(\cdot)$ denotes the collaborative weights generator with parameters $\Theta$.

The only module that requires training is the collaborative weights generator, which converts pre-trained collaborative features to collaborative weights. We minimize prediction errors to optimize the parameters of the generator $\Theta$:
\begin{equation}
    \hat{\Theta} = argmin_{\Theta}\sum_{(u,i,y)\in\mathcal{D}}\ell(\hat{y},y),
\end{equation}
where $\ell(\cdot)$ denotes the loss function, which is implemented as the binary cross-entropy (BCE) loss.

\section{Experiments}

\subsection{Experimental Settings}

\begin{table}[tbp]
    \centering
    \scalebox{0.85}{
    \begin{tabular}{cccccc}
     \toprule
        Dataset &  \#Train & \#Valid & \#Test & \#User & \#Item \\
        \midrule
        ML-1M &  33,891 & 10,401 & 7,331 & 839 & 3,256 \\
        Amazon-Book &  727,468 & 25,747 & 25,747 & 22,967 & 34,154 \\
        \bottomrule
    \end{tabular}
    }
    \caption{Statistics of the processed datasets.}
    \label{tab:datasets}
\end{table}

\begin{table*}[ht]
    \centering
    \begin{adjustbox}{max width=\textwidth}
    \begin{tabular}{cc|cccccc}
        \toprule
        \multicolumn{2}{c|}{\textbf{Dataset}} & \multicolumn{3}{c}{Amazon-Book} & \multicolumn{3}{c}{ML-1M} \\
        \cmidrule(lr){1-2} \cmidrule(lr){3-5} \cmidrule(lr){6-8}
        \multicolumn{2}{c|}{\textbf{Method}} & AUC & UAUC & Improve & AUC & UAUC & Improve \\
        \midrule
        \multirow{3}{*}{\textbf{Collab.}} 
        & MF & 0.7105 & 0.5543 & 14.04\% & 0.6486 & 0.6396 & 10.56\% \\
        & LightGCN & 0.7026 & 0.5619 & 13.93\% & 0.5858 & 0.6512 & 15.68\% \\
        & SASRec & 0.6675 & 0.5614 & 17.04\% & 0.7005 & 0.6734 & 3.65\% \\
        \midrule
        \multirow{3}{*}{\textbf{LLMRec}} 
        & ICL & 0.5180 & 0.5043 & 51.61\% & 0.5119 & 0.5178 & 38.37\% \\
        & Prompt4NR & 0.6527 & 0.5011 & 25.10\% & 0.7027 & 0.6713 & 3.28\% \\
        & TALLRec & 0.6583 & 0.4971 & 25.11\% & 0.7044 & 0.6741 & 3.31\% \\
        \midrule
        \multirow{5}{*}{\textbf{\makecell{LLMRec\\ w/ Collab.}}} 
        & PersonPrompt & 0.7113 & 0.5596 & 13.44\% & 0.7014 & 0.6503 & 5.40\% \\
        & CoLLM-MF & 0.8021 & 0.5782 & 5.14\% & 0.7028 & 0.6714 & 3.64\% \\
        & CoLLM-LGCN & 0.7835 & 0.5663 & 7.48\% & 0.7164 & 0.6842 & 4.68\% \\
        & CoLLM-SAS & 0.7538 & 0.5874 & 7.55\% & 0.7059 & 0.6531 & 4.84\% \\
        & BinLLM & 0.8157 & 0.5724 & 4.83\% & 0.7132 & 0.6815 & 2.11\% \\
        \midrule
        \multirow{3}{*}{\textbf{Ours}} 
        & CoRA-MF & \textbf{0.8179} & \textbf{0.6262} & - & \textbf{0.7361} & \textbf{0.6884} & - \\
        & CoRA-LGCN & 0.7886 & 0.5689 & - & 0.7128 & 0.6966 & -  \\
        & CoRA-SAS & 0.7677 & 0.5961 & - & 0.7019 & 0.6517 & - \\
        \bottomrule
    \end{tabular}
    \end{adjustbox}
    \caption{Performance comparison of various models on Amazon-Book and ML-1M. "Collab." denotes collaborative recommendation methods. "Improve" denotes the relative improvement of CoRA compared to baselines, averaged over the two metrics. All improvements are statistically significant, as determined by a paired t-test with $p \leq 0.05$.}
    \label{tab:overall}
\end{table*}

\noindent \textbf{Datasets.} We adopt two widely-used public datasets for evaluation: ML-1M\footnote{\url{https://grouplens.org/datasets/movielens/1m/}} and Amazon-Book\footnote{\url{https://jmcauley.ucsd.edu/data/amazon/index_2014.html}}. For dataset processing, we adhere entirely to the setup of CoLLM~\cite{CoLLM}. The detailed statistics of the datasets are summarized in Tab.~\ref{tab:datasets}.

\ \\ \noindent \textbf{Baselines.} To assess the effectiveness of CoRA, we compare it with three types of methods: conventional collaborative filtering methods (MF~\cite{bpr}, LightGCN~\cite{lightgcn}, and SASRec~\cite{sasrec}), LLMRec methods without collaborative information (ICL~\cite{icl}, Prompt4NR~\cite{prompt4nr}, and TALLRec~\cite{TALLRec}), and LLMRec methods that consider collaborative information(PersonPrompt~\cite{TOIS23-PEPLER}, CoLLM~\cite{CoLLM}, and BinLLM~\cite{Textlike}).

\ \\ \noindent \textbf{Implementation Details.} We use Vicuna-7B as the backbone LLM to implement our method. Binary Cross-Entropy (BCE) is employed as the objective function for optimizing all methods. AdamW~\cite{Loshchilov2017DecoupledWD} optimizer is adopted for optimizing LLM, and Adam~\cite{Kingma2014AdamAM} optimizer for other methods. The detailed hyper-parameter settings and exploration can be found in the Appendix and the source code. We tune hyper-parameters according to the AUC metric on the validation dataset.

Regarding the collaborative weights generator, we set the hidden dimension and the embedding size of the collaborative model $d_c$ as 256 and the number of blocks $N$ as 8. The rank $r$ of perceptual weights is 16. The number of perceptual queries $k$ is 4. We insert collaborative weights $\Delta W$ on every decoder block of LLM. For better collaborative perceptual ability, we explore to insert $\Delta W$ for different types of weights in LLM, including $[W_Q,W_K,W_V,W_O,W_{FFN}]$. It is worth noting that the last linear layer $W_{proj}$ of the collaborative weights generator is zero-initialized for training stability as it is equivalent to the $\Delta W_B$ of LoRA weights.

For our CoRA, we implement it across three collaborative filtering methods to validate the effectiveness of our method in utilizing different types of collaborative information, denoted as CoRA-MF, CoRA-SASRec, and CoRA-LightGCN.

\ \\ \noindent \textbf{Evaluation Metrics.} We employ two widely used metrics for click/rating prediction: AUC (Area under the ROC Curve, which measures the overall prediction accuracy) and UAUC (AUC averaged over users)~\cite{Liu2021ConceptAwareDG} to evaluate the performance of our method and baselines.

\ \\ \noindent \textbf{Hyperparameter Settings.} Regarding hyperparameter tuning, we explore the learning rate within the range of $[1e-2, 1e-3, 1e-4]$ and the weight decay within the range of $[1e-2,\dots,1e-6]$. Finally, we train the collaborative weights generator with a learning rate of $1e-2$ on Amazon-Book and $1e-3$ on ML-1M. On two datasets, the warm-up learning rate is set to $1e-5$. The rank $r$ of the collaborative weights is set to 16, and we insert collaborative weights into the weights of four types of LLMs, namely $[W_Q,W_K,W_V,W_O]$ after comparing the performances. For all pre-training collaborative filtering models, we set the embedding size to $256$. The learning rate is explored within the range of $[1e-1,\dots,1e-5]$, and the weight decay is explored within the range of $[1e-2,\dots,1e-6]$. 

Besides, we adopt an early-stop strategy if AUC on the validation set no longer increases after $20$ epochs to avoid overfitting and achieve the best performance. Our source code included in the Supplementary material provides detailed hyper-parameter settings.

\begin{figure}[t]
    \centering
    \subfigure[Amazon-Book Warm]{
        \includegraphics[width=0.45\linewidth]{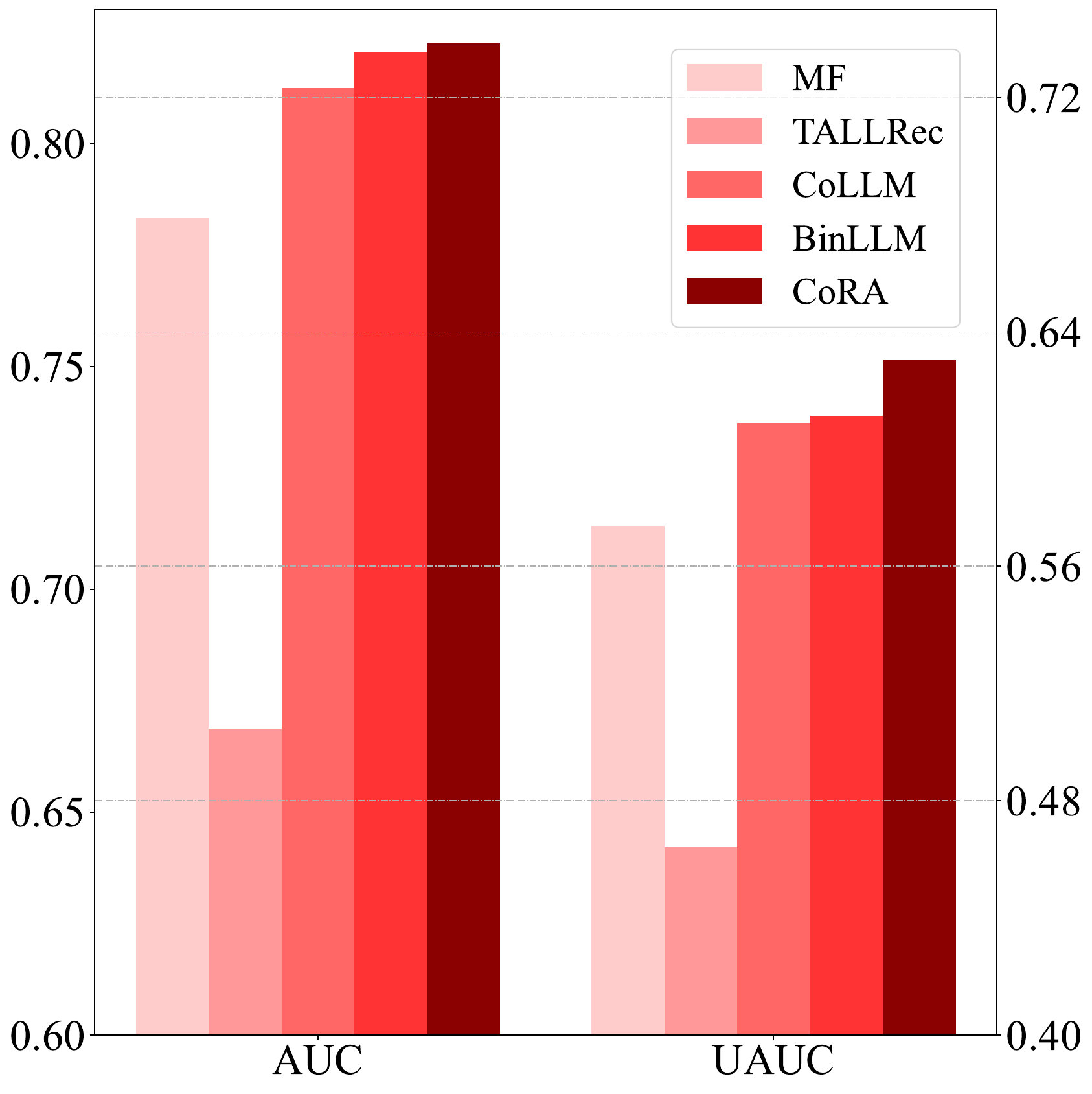}
    }
    \subfigure[ML-1M Warm]{
        \includegraphics[width=0.45\linewidth]{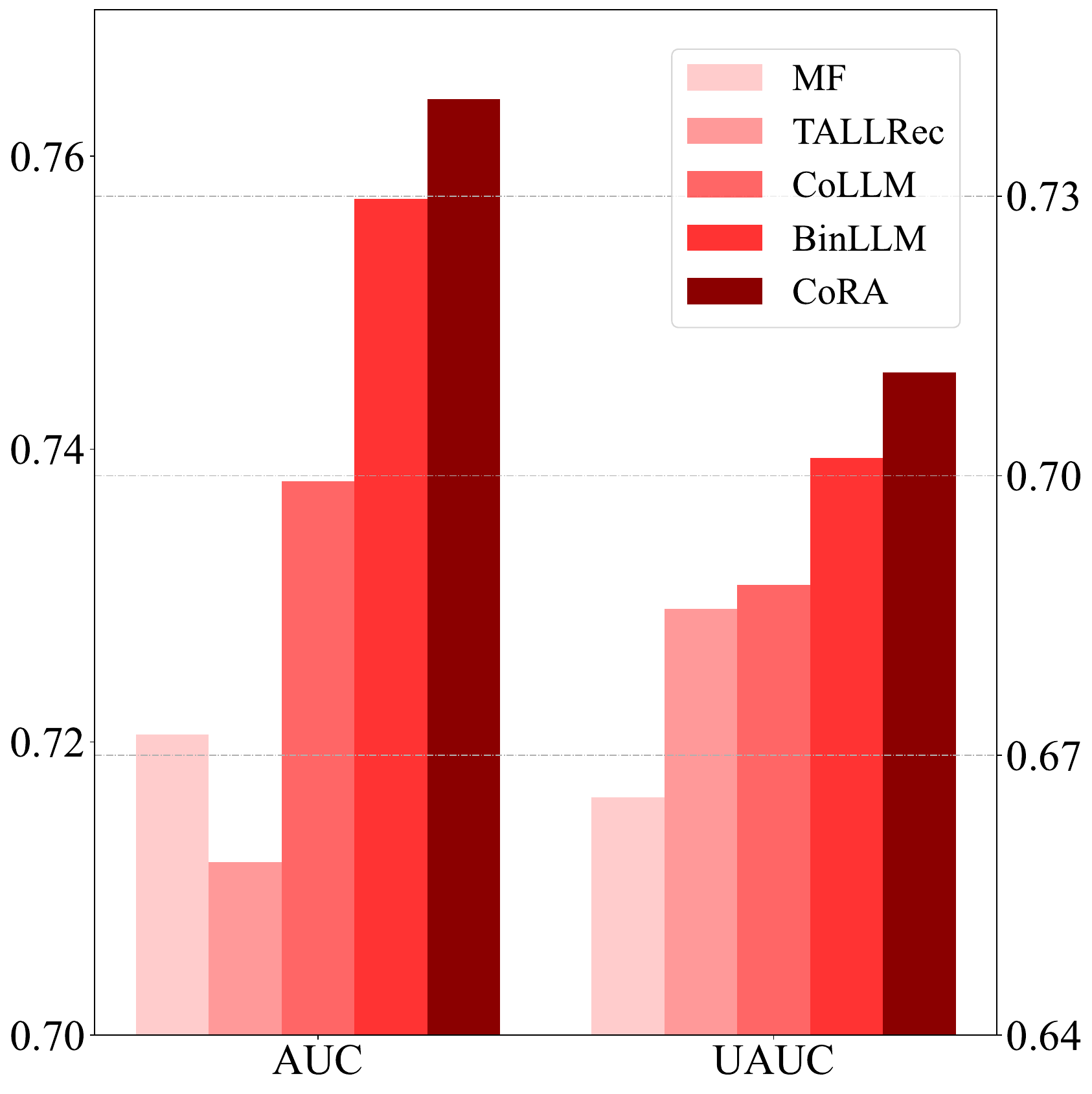}
    }
    \subfigure[Amazon-Book Cold]{
        \includegraphics[width=0.45\linewidth]{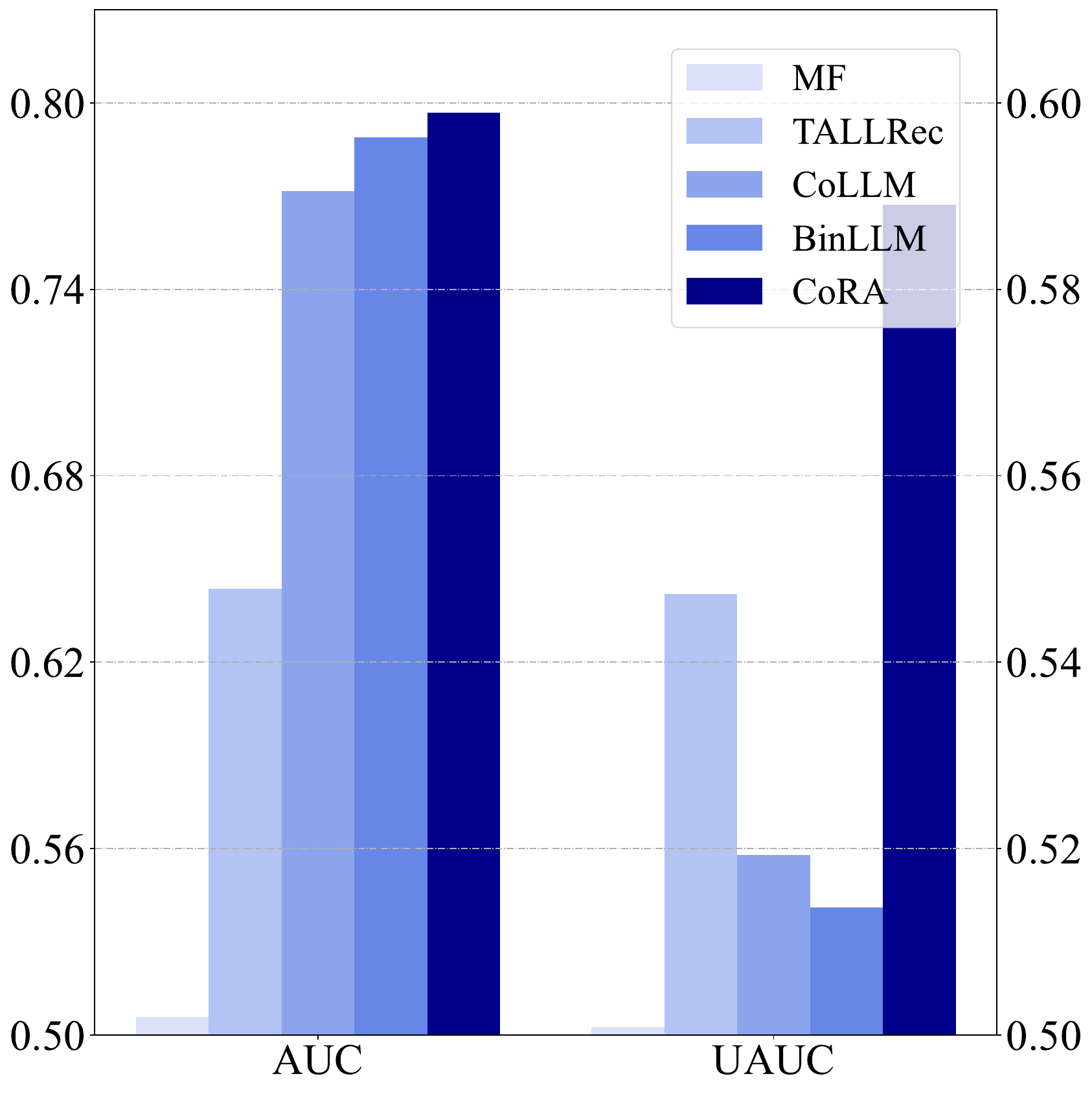}
    }
    \subfigure[ML-1M Cold]{
        \includegraphics[width=0.45\linewidth]{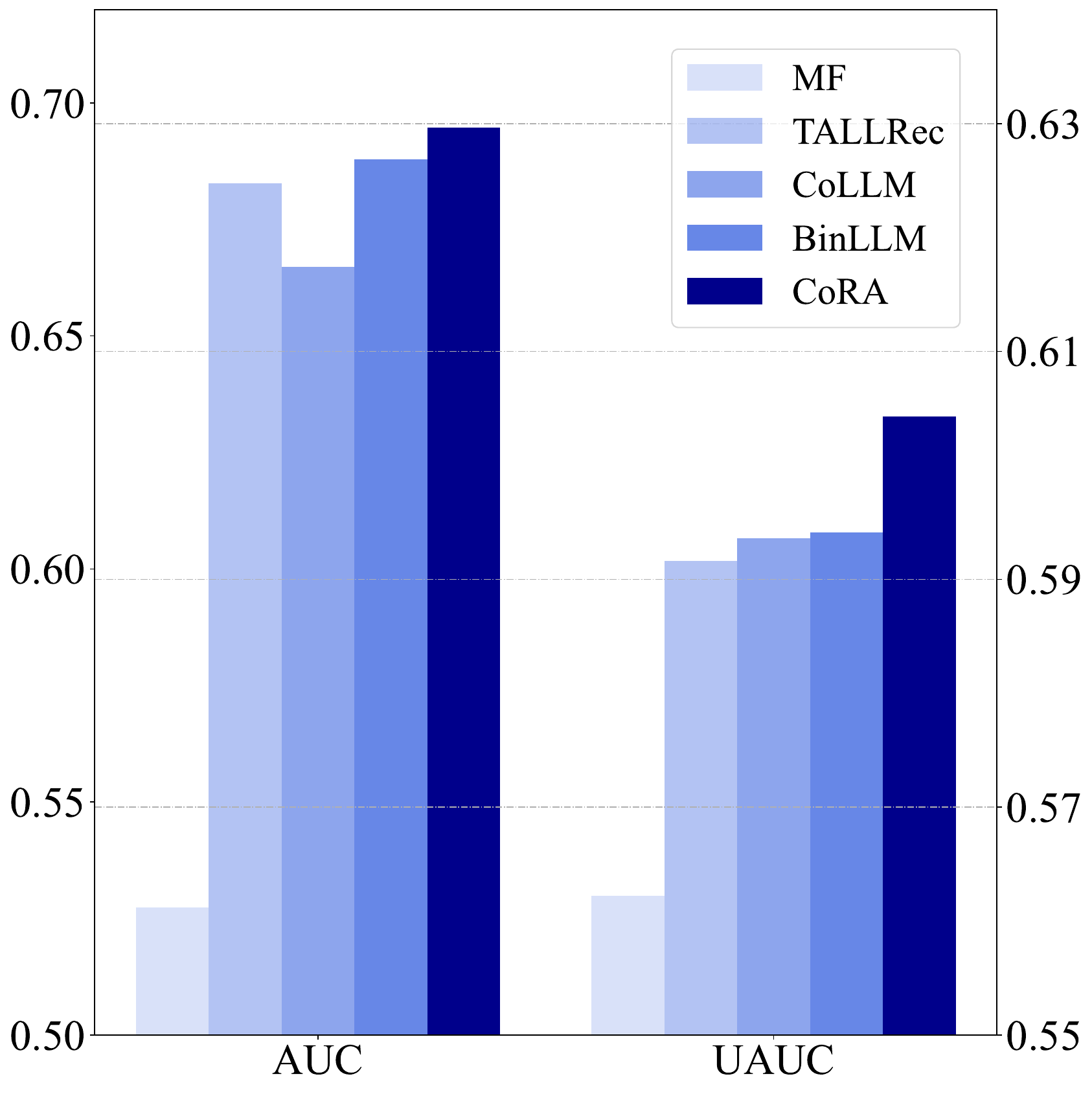}
    }
    \caption{Performance comparison in warm and cold scenarios on Amazon-Book and ML-1M. The left and right y-axis are AUC and UAUC, respectively.}
    \label{fig:warm-cold}
\end{figure}

\subsection{Performance Comparison}

\noindent \textbf{Overall Performance.}
The performance is summarized in Tab.~\ref{tab:overall}, from which we can find that our CoRA outperforms both collaborative filtering and LLMRec baselines, confirming the superiority of CoRA in leveraging collaborative information from traditional CF methods and general knowledge from LLM to make better predictions. 

LLMRec with collaborative information surpasses traditional collaborative filtering methods, revealing that LLM's extensive knowledge can be effectively generalized to recommendation systems. Furthermore, LLMRec methods that incorporate collaborative information show superior performance compared to those that do not, highlighting the significance of incorporating collaborative data into LLM for enhanced recommendations. Overall, integrating collaborative information with LLM's general abilities represents a promising advancement in recommendation technology.

Our approach retains its advantages compared to LLMRec with collaborative information methods. When comparing the variants of CoRA and CoLLM based on different collaborative filtering models, CoRA consistently beats CoLLM. This indicates that aligning collaborative information with LLM's parameter space is more effective in helping LLM perceive collaborative information and integrate it with linguistic competence.

\ \\ \noindent \textbf{Warm and Cold Performance.}
The main purpose of introducing collaborative information into LLMRec is to enhance the performance of LLMRec methods in warm-start scenarios. We divide the test data into warm and cold subsets based on the number of interactions. Without loss of generality, we compare the testing performance of MF, TALLRec, CoLLM, BinLLM, and CoRA on both warm and cold test subsets, as shown in Fig.~\ref{fig:warm-cold}.

In the warm scenario, TALLRec performs worse than MF because it does not consider collaborative information. Conversely, LLMRec methods that integrate collaborative information perform better, with CoRA achieving the best. This demonstrates that collaborative information is crucial for recommendations in warm-start scenarios, and our method is more effective in utilizing it.

In the cold start scenario, all LLMRec methods outperform MF, indicating that LLM's universal capabilities can effectively alleviate the cold start problem by utilizing item textual information. Moreover, CoRA enhances the cold-start performance, suggesting that our method is superior to existing methods in integrating the collaborative knowledge of CF models and the general language knowledge of LLM.

\subsection{Ablation Study}

\begin{figure}[t]
    \centering
    \subfigure[Amazon]{
        \includegraphics[width=0.45\linewidth]{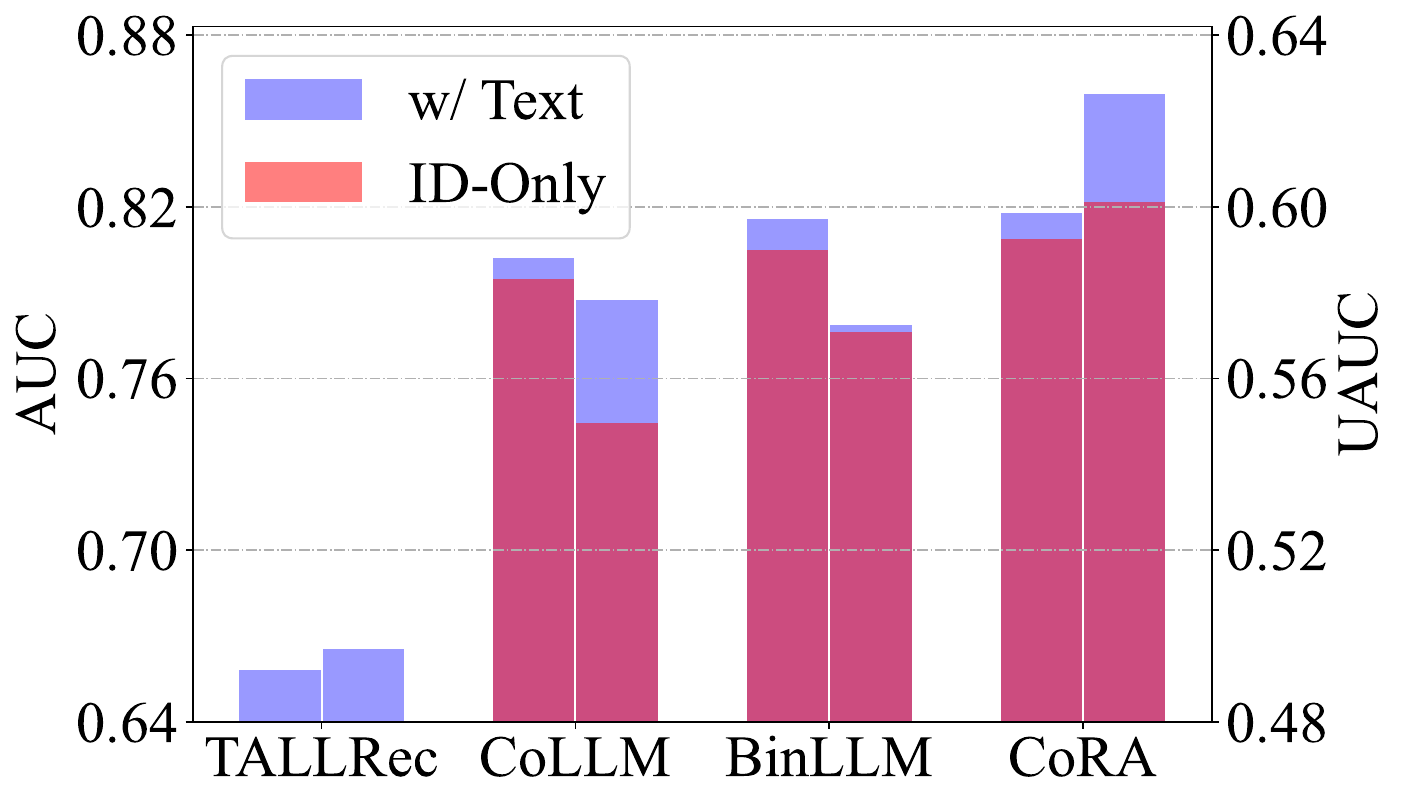}
    }
    \subfigure[ML-1M]{
        \includegraphics[width=0.45\linewidth]{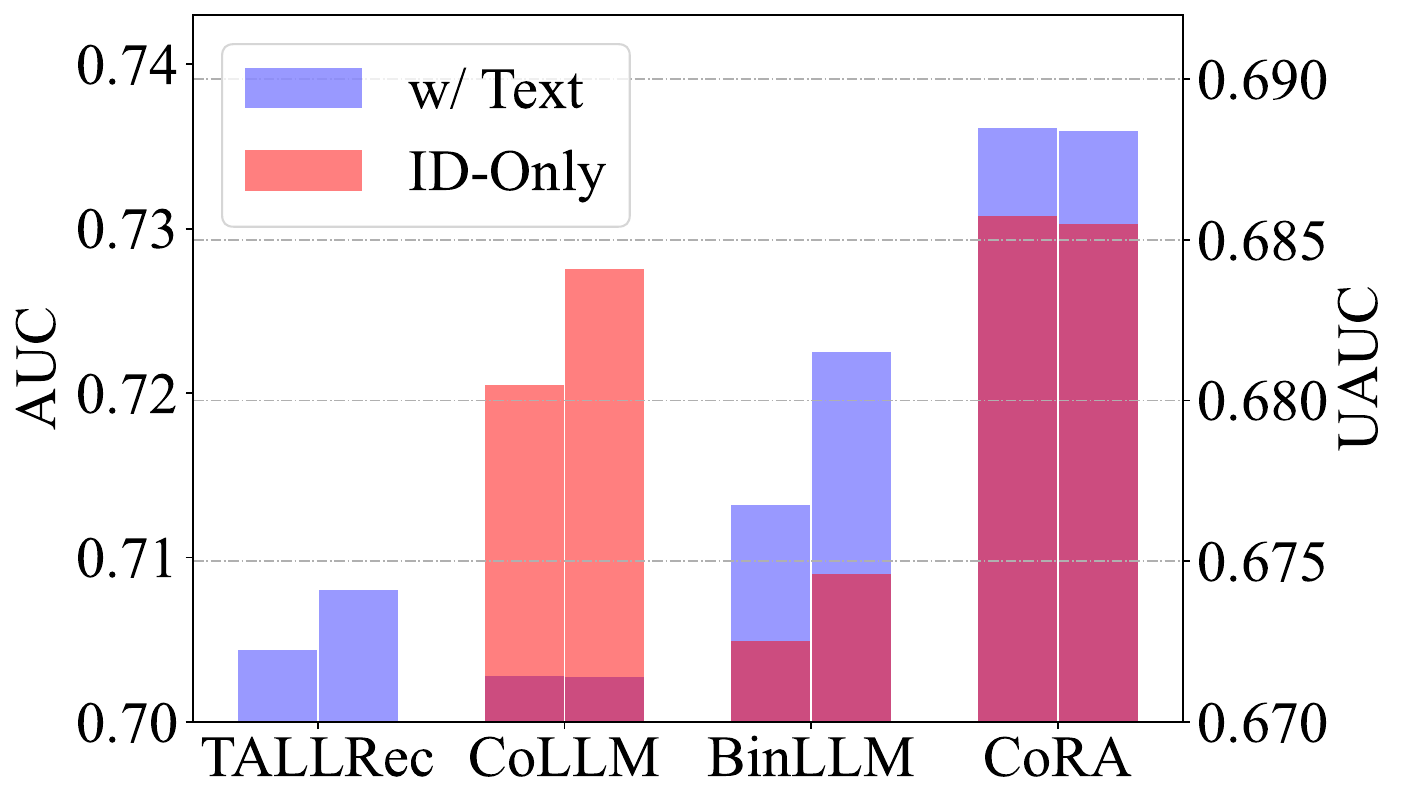}
    }
    \caption{Performance of various variants. ``ID-Only" refers to the removal of the item text. ``w/ Text" represents adding item textual descriptions.}
    \label{fig:idonly}
\end{figure}

\noindent \textbf{The effectiveness of integrating collaborative and textual information.}
To further validate the superiority of our method in enabling LLM to incorporate collaborative information, we construct variants of different LLMRec methods only with ID embeddings (i.e., removing textual descriptions of items). We treat TALLRec as a ``Text-Only" variant of the three methods as a reference. The results are shown in Fig.~\ref{fig:idonly}, from which we can observe that our method outperforms the baselines when using only ID embeddings, confirming that our method can significantly enhance the collaborative perception ability of LLM. On this basis, the performance of all methods improved after involving item textual descriptions in most cases. Notably, the performance of CoLLM on ML-1M significantly declined after introducing textual information, indicating that aligning ID embeddings with LLM's input space interferes with textual semantics and fails to combine their respective advantages. Our CoRA, on the other hand, shows a greater enhancement after incorporating text information, which we attribute to introducing collaborative information in LLM's parameter space instead of the input space, fundamentally avoiding the interference problem.

\begin{table}[t]
    \centering
    \begin{adjustbox}{max width=\textwidth}
    \begin{tabular}{ccccc}
    \toprule
        \multicolumn{1}{c}{\multirow{2}{*}{\textbf{Weight Type}}} & \multicolumn{2}{c}{Amazon-Book} & \multicolumn{2}{c}{ML-1M} \\
        \cmidrule(lr){2-3} \cmidrule(lr){4-5}
        & AUC & UAUC & AUC & UAUC \\
        \midrule
        qkvof & 0.8141 & 0.6068 & 0.7312 & 0.6801 \\
        qkvo & \textbf{0.8179} & \textbf{0.6262} & \textbf{0.7361} & \textbf{0.6884} \\
        qkv & 0.7741 & 0.5747 & 0.6947 & 0.5933 \\
        qko & 0.8091 & 0.5949 & 0.7111 & 0.5973 \\
        qk & 0.7685 & 0.5644 & 0.6784 & 0.5887 \\
        \bottomrule
    \end{tabular}
    \end{adjustbox}
    \caption{The impact of perceptual weighs type. q, k, v, and o denote query, key, value, and output linear weights in the self-attention module, respectively. f denotes the weights of feed-forward networks.}
    \label{tab:weight}
\end{table}

\ \\ \noindent \textbf{The type of equipped collaborative weights.}
As mentioned in \textbf{preliminaries}, there are six types of weights in an LLM's decoder block, which are \textbf{q}uery, \textbf{k}ey, \textbf{v}alue, \textbf{o}utput, and up\&down (\textbf{f}eed-forward). We explore the impact of inserting collaborative weights for different types of LLM weights. As shown in Tab.~\ref{tab:weight}, LLM equipped with collaborative weights for all types except feed-forward of weights performs best. Moreover, we observe that the performance of \emph{qkvo} and \emph{qko} is much better than \emph{qkv}, suggesting that the output weights are essential for collaborative perception.

\section{Conclusion}

In this paper, we first explore the issues arising from aligning collaborative information in the input space of LLMs. To address them, we introduce CoRA, a novel paradigm that enables LLM to perceive collaborative information without fine-tuning or extra collaborative tokens. Our method converts collaborative information into LLM's incremental weights through a collaborative weights generator, effectively integrating collaborative and textual information. Extensive experiments demonstrate the superiority of CoRA. For future work, we will expand experiments on other LLM backbones and recommendation tasks. Besides, we aim to extend our method to device-cloud collaborative learning.

\bibliography{aaai25}

\newpage

\section{Details for Baselines}

The baselines consist of three types of methods: conventional collaborative filtering methods (MF, LightGCN, and SASRec), LLMRec methods without collaborative information (ICL, Prompt4NR, and TALLRec), and LLMRec methods that consider collaborative information(PersonPrompt, CoLLM, and BinLLM).

\subsection{Collaborative Filtering Models}
\begin{itemize}
    \item \textbf{MF} \cite{bpr}: This model is a classic item ranking method built upon the assumption that a user prefers an interacted item to an unknown one.
    \item \textbf{LightGCN} \cite{lightgcn}: This model abandons the use of feature transformation and nonlinear activation, and only retains the most important neighbor aggregation module in GCNs for collaborative filtering. 
    \item \textbf{SASRec} \cite{sasrec}: It adopts the multi-head self-attention mechanism to perform sequential recommendations.
\end{itemize}

\subsection{LLMRec without Collaborative Information}
\begin{itemize}
    \item \textbf{ICL} \cite{icl}: This is an In-Context Learning method for LLMRec, which directly utilizes frozen LLM for recommendations.
    \item \textbf{Prompt4NR} \cite{prompt4nr}: This is a state-of-the-art LLMRec method of using soft prompt paradigm. We extend it to LLMs, taking the implementation in CoLLM.
    \item \textbf{TALLRec} \cite{TALLRec}: This is a state-of-the-art LLMRec method that aligns LLMs with recommendation tasks through instruct tuning.
\end{itemize}

\subsection{LLMRec with Collaborative Information}
\begin{itemize}
    \item \textbf{PersonPrompt} \cite{TOIS23-PEPLER}: This LLMRec method integrates collaborative information by inserting new tokens to represent users and items.
    \item \textbf{CoLLM} \cite{CoLLM}:  It integrates collaborative information by mapping user and item embedding into the input space of LLMs.
    \item \textbf{BinLLM} \cite{Textlike}:  This method converts collaborative embeddings from external models into binary sequences that LLMs can understand and operate on directly.

\end{itemize}

\section{Details for Implementations}

\subsection{Computing Infrastructure}

All experiments are performed on a single NVIDIA A100 80GB GPU with an AMD EPYC 7763 64-Core CPU and 440GB RAM. All models are implemented using the PyTorch framework version 2.3.1+cu113.

\subsection{Pilot Study}

\begin{table}
\centering
    \begin{tabular}{p{7.5cm}} 
    \toprule
    \{``input": ``Please summarize the user's preferences, the user has given high ratings to the following books: Exodus (The New Frontiers Series),The Intern (The Forbidden World), Freedom's Fury (Freedom's Fire) (Volume 2).", ``ground\_truth": ``The user seems to enjoy science fiction books with themes of adventure, rebellion, and freedom. They appreciate fast-paced plots and well-developed characters."\}\\
    \bottomrule
    \end{tabular}
    \caption{Example of UProfile dataset, GPT-3.5-Turbo generates the ground truth.}
    \label{tab:profile}
\end{table}

\begin{table}
\centering
    \begin{tabular}{p{7.5cm}} 
    \toprule
    \{``input": ``Given the following nine book titles, which one is the most relevant to the target book title ``Rosie's Song"? Please select the most appropriate word. Options: 1. Proust and the Sense of Time, 2. Seeker of Horizons, 3. Perfect Passwords: Selection, Protection, Authentication, 4. Partnership with Christ: A Cistercian Retreat (Monastic Wisdom series), 5. Angels of Death: Inside the Biker Gangs' Crime Empire, 6. LA Shorts, 7. Jesus Symbol of God, 8. Follow You, 9. Word Clues the Vocabulary Buil. Please respond with only one book title.", ``ground\_truth": ``Follow You."\}\\
    \bottomrule
    \end{tabular}
    \caption{Example of ITMatching dataset, GPT-3.5-Turbo generates the ground truth.}
    \label{tab:title}
\end{table}

In Sec.\textbf{Introduction}, we evaluate the LLM's interpreting and inferring capabilities for general and recommendation-related tasks before and after fine-tuning. We fine-tune Vicuna-7B on \emph{Amazon-Book} using the prompt in TALLRec.

For general ability, we follow \cite{LLMSurvey}\footnote{\url{https://github.com/RUCAIBox/LLMSurvey/}} to evaluate the LLM in the datasets \emph{WikiFact}~\cite{wikifact}, \emph{SocialIQA}~\cite{socialiqa} and \emph{XSum}~\cite{XSum}. We adopt the metrics Exact-Match (EM) scores of generated answers on \emph{WikiFact} and \emph{SocialIQA}, and ROUGE-L scores on \emph{XSum}.


For recommendation ability, we construct UProfile and ITMatching datasets based on \emph{Amazon-Book} for user profiling and item title matching, respectively. To construct UProfile dataset, we select users with interactions greater than two and less than ten in \emph{Amazon-Book}. We then organize the historical interaction items for each user as a prompt and use GPT-3.5-Turbo to summarize the user preference as the ground truth. Table~\ref{tab:profile} shows an example of organized data. To construct the ITMatching dataset, we randomly sampled 1000 item title sets with a size of 10. We randomly select a target title for each title set and request GPT-3.5-Turbo to find the most relevant title in this set as the ground truth. An example in the ITMatching dataset is given in Table~\ref{tab:title}.

Finally, we utilize Vicuna-7B before and after fine-tuning with the TALLRec method to summarize user preferences and match item titles. Then, we calculate ROUGE-L scores to evaluate LLM's user profiling and item title understanding capability. Please refer to the source code in the Supplementary Material for detailed implementation.

\end{document}